\newif\ifreviewmode
\newif\iflinenumbers

\RequirePackage{xcolor} 
\documentclass[journal,twoside,web]{ieeecolor}

\usepackage{etoolbox}
\makeatletter
\@ifundefined{color@begingroup}%
  {\let\color@begingroup\relax
   \let\color@endgroup\relax}{}%
\def\fix@ieeecolor@hbox#1{%
  \hbox{\color@begingroup#1\color@endgroup}}
\patchcmd\@makecaption{\hbox}{\fix@ieeecolor@hbox}{}{\FAILED}
\patchcmd\@makecaption{\hbox}{\fix@ieeecolor@hbox}{}{\FAILED}

\definecolor{subsectioncolor}{rgb}{0,0,0} 
\makeatletter\@namedef{ver@fixltx2e.sty}{}\makeatother
\usepackage{dblfloatfix}
\usepackage{placeins}
\usepackage{xspace}
\usepackage[per-mode=symbol]{siunitx}
\usepackage{amsmath,amssymb,amsfonts}
\usepackage{graphicx}
\usepackage{hyperref}
\hypersetup{hidelinks}
\usepackage{textcomp}
\usepackage{bm}
\usepackage{booktabs}
\makeatletter\def\caption@documentclass{standard}\makeatother
\usepackage{caption}
\usepackage{tabularx}
\usepackage{array}
\usepackage[table]{xcolor}
\usepackage[edges]{forest}
\usepackage[T1]{fontenc}
\usepackage{tikz}
\usetikzlibrary{positioning,calc,fit,arrows.meta,decorations.pathreplacing,backgrounds}
\usepackage{helvet}

\makeatletter
\def\ps@titlepagestyle{\def\@oddfoot{}\def\@evenfoot{}%
   \def\@oddhead{\begin{tabular}{@{}p{43pc}@{}}\\[-24pt]\vbox{\hsize43pc\scriptsize\textsf\leftmark \hfill \textsf\thepage\hbox{}}\\[-19pt]
         \vbox{\color{subsectioncolor}\hrule height1pt width43pc depth0pt} \end{tabular}}%
   \let\@evenhead\@oddhead}
\makeatother

\definecolor{tabrow}{HTML}{F2F5F7}

\usepackage[nolist,nohyperlinks,printonlyreused]{acronym}
\begin{acronym}
\acro{DM}{diffusion model}
\acro{MRI}{magnetic resonance imaging}
\acro{CT}{computed tomography}
\acro{CS}{compressed sensing}
\acro{DL}{deep learning}
\acro{UFF}{Ultrasound File Format}
\acro{USTB}{UltraSound ToolBox}
\acro{RF}{radio-frequency}
\acro{RFP}{request for proposals}
\acro{IQ}{in-phase and quadrature}
\acro{IVUS}{intravascular ultrasound}
\acro{ULM}{ultrasound localization microscopy}
\acro{ARFI}{acoustic radiation force impulse}
\acro{CMPE}{common midpoint phase error}
\acro{LOC}{lag-one coherence}
\acro{CNR}{contrast-to-noise ratio}
\acro{GPU}{graphics processing unit}
\acro{SoS}{speed of sound}
\acro{DAS}{delay-and-sum}
\acro{SSA}{swept synthetic aperture}
\acro{USCT}{ultrasound computed tomography}
\end{acronym}

\usepackage[
    style=ieee,            
    doi=false,             
    isbn=false,            
    url=false,              
    eprint=true,           
    maxbibnames=6,         
    minbibnames=1,         
    giveninits=true,       
    sorting=none,          
    backend=biber
]{biblatex}

\ifreviewmode
\definecolor{wessel}{HTML}{129f57}
\definecolor{tristan}{HTML}{1231f1}
\definecolor{simon}{HTML}{00008B}
\definecolor{oisin}{HTML}{c94520}
\definecolor{felix}{HTML}{FF00FF}
\else
\colorlet{wessel}{black}
\colorlet{tristan}{black}
\colorlet{simon}{black}
\colorlet{oisin}{black}
\colorlet{felix}{black}
\fi

\newcommand{\zea}{\texttt{zea}\xspace}
\DeclareRobustCommand{\OpenHRF}{Open\nobreakdash-H\nobreakdash-RF\xspace}
\newcommand{\OpenHRFSize}{\qty{40}{\tera\byte}\xspace}
\newcommand{\OpenHRFAcquisitions}{\num{19332}\xspace}
\newcommand{\OpenHRFInstitutions}{\num{39}\xspace}
\newcommand{\OpenHRFContributors}{\num{34}\xspace}
\newcommand{\OpenHRFFrames}{\num{25092732}\xspace}
\newcommand{\OpenHRFTransmitEvents}{\num{241078230}\xspace}

\usepackage{amsmath,amsfonts,bm}

\def\1{\bm{1}}

\DeclareMathAlphabet{\mathsfit}{\encodingdefault}{\sfdefault}{m}{sl}
\SetMathAlphabet{\mathsfit}{bold}{\encodingdefault}{\sfdefault}{bx}{n}

\definecolor{zblue}{HTML}{1F5C8B}   
\definecolor{zred}{HTML}{A8452C}    
\definecolor{zgray}{HTML}{6B7580}   
\definecolor{zlight}{HTML}{B9C0C7}  
\definecolor{zfill}{HTML}{F1F3F5}   
\definecolor{zamber}{HTML}{E8A33D}  
 
\tikzset{
  zbox/.style   ={rounded corners=2pt, draw=zlight, fill=zfill, line width=0.5pt,
                  inner xsep=4pt, inner ysep=3pt, font=\ttfamily\scriptsize},
  zcall/.style  ={zbox, draw=zamber!80!black, fill=zamber!18},
  zarrow/.style ={-{Stealth[length=3.5pt,width=2.6pt]}, draw=zgray, line width=0.6pt},
  zline/.style  ={draw=zgray, line width=0.6pt},
  zstep/.style  ={circle, draw=zgray, fill=white, inner sep=0pt, minimum size=8pt,
                  font=\sffamily\tiny, text=zgray},
  zlab/.style   ={font=\sffamily\scriptsize, text=zgray},
  zframe/.style ={rounded corners=2pt, dashed, dash pattern=on 2pt off 1.5pt, line width=0.5pt},
}

\iflinenumbers
\usepackage[switch,pagewise]{lineno}

\linenumbers
\fi

\begin{document}
\title{Open-H-RF: A Large-Scale Ultrasound\\Raw Channel Dataset Initiative
{
\thanks{The steering committee of the Open-H-RF consortium consists of NVIDIA, Eindhoven University of Technology (TU/e) and Stanford University. \authorcontributions}
}
}

\newcommand{\authorcontributions}{%
  \textsuperscript{*}Equal contribution as first authors.
  \textsuperscript{\dag}Equal contribution as senior authors.
  \textsuperscript{\S}Core team.%
}

\author{%
\mbox{T.S.W. Stevens\textsuperscript{*,\S}}, \mbox{S.W. Penninga\textsuperscript{*,\S}}, \mbox{W.L. van Nierop\textsuperscript{*,\S}}, \mbox{O. Nolan\textsuperscript{*,\S}}, \mbox{F. Duelmer\textsuperscript{\S}}, \mbox{B.N. Frey\textsuperscript{\S}},
\\[.3ex]
\mbox{B. Federici}, \mbox{V. van Hal}, \mbox{L.D. van Harten}, \mbox{H. de Hoop}
% ...
}

\makeatletter\g@addto@macro\@author{\par\vspace{-14pt}}\makeatother

\maketitle

\begin{figure*}
    \centering
    \includegraphics[width=0.98\linewidth]{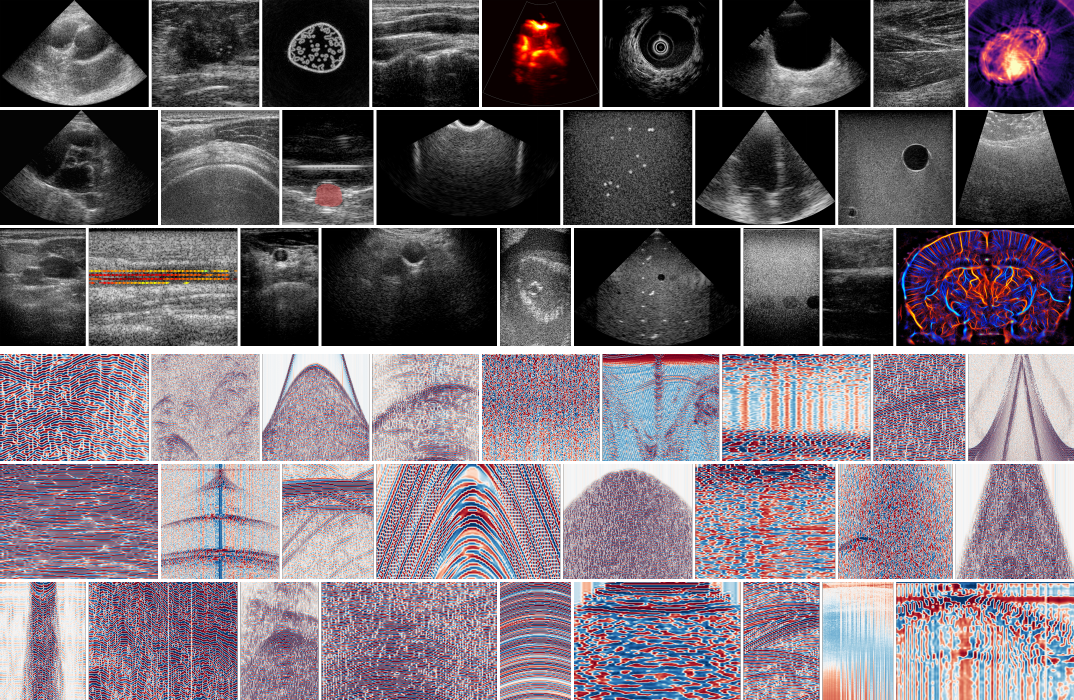}
    \caption{Impression of the dataset. \textit{Top:} reconstructed images from a selection of samples across the contributions. \textit{Bottom:} the corresponding (demodulated) raw channel data from the same samples, in matching order. Because the full raw channel data required for reconstruction is a far larger tensor (spanning all transmits, receive elements, and axial samples) that cannot be displayed directly, each \ac{RF} panel shows only a single, randomly selected transmit, cropped for visualization, and is meant as an impression of the underlying signal. The dataset itself stores the complete raw channel data, alongside all parameters and metadata needed to derive the reconstructions.}
    \label{fig:collage}
\end{figure*}

\begin{abstract}
\sisetup{mode=text, reset-text-family=false, reset-text-series=false}
Medical ultrasound is a widely accessible, real-time imaging modality, yet progress in data-driven ultrasound research is constrained by the scarcity of large, open collections of raw channel data. Existing public datasets are predominantly based on reconstructed images, while the smaller number of open raw channel datasets is generally specialized to individual reconstruction or imaging tasks, limiting the development of general-purpose methods and foundation models that operate directly on ultrasound signals. While interpretive tasks can be performed on reconstructed images, many of the key technical challenges in ultrasound imaging, such as aberration correction and reverberation reduction, require the raw channel data itself. We introduce \textit{\OpenHRF}, a community-driven open dataset initiative comprising approximately \OpenHRFSize of medical ultrasound raw channel data in \OpenHRFAcquisitions acquisitions, contributed by \OpenHRFInstitutions institutions across academia and industry, spanning \emph{in-vivo}, phantom, and simulated data. Each dataset is annotated with a number of supported tasks, ranging from generalized reconstruction, speed-of-sound estimation, and aberration correction to blood flow imaging, motion estimation, quantitative ultrasound, and image interpretation. \OpenHRF provides a unified, self-contained data format that captures raw channel data together with the acquisition, probe, metadata, and reconstruction information required for reproducible processing across scanners and acquisition schemes. The corpus is publicly hosted under a permissive CC BY 4.0 license and can be streamed programmatically and selectively at scale. By bringing heterogeneous ultrasound data into a common representation, backed by modern data infrastructure, \OpenHRF lowers the barrier to working with raw channel data and aims to open medical ultrasound research to new communities. See \url{https://open-h.github.io/Open-H-RF}.
\end{abstract}


\section{Introduction}
\label{sec:introduction}

\IEEEPARstart{O}{pen} datasets have repeatedly acted as catalysts for progress in data-driven research. They provide common benchmarks and standardized baselines against which new methods can be compared fairly, form the basis of community challenges, and enable scientific discovery. In computer vision, large open collections of natural images have driven progress for over a decade~\cite{dengImageNetLargescaleHierarchical2009}. Similarly, in medical imaging, open datasets have enabled progress across modalities. In \ac{MRI}, for instance, the \textit{fastMRI} dataset~\cite{knoll2020fastmri}, consisting of thousands of raw \textit{k-space} samples, has spurred the development of dozens of novel approaches to \ac{MRI} reconstruction~\cite{sriram2020end, pineda2020active}. Large, standardized, openly shared datasets have also become the key enabler of foundation models: models pretrained on large and diverse corpora whose weights can be fine-tuned for a wide range of downstream tasks~\cite{bommasani2021foundation}. In medical imaging, such models are now trained on large collections of \ac{CT} and \ac{MRI} data~\cite{pai2025ctfm,tak2026brainiac}. In ultrasound imaging, USFM was pretrained on more than two million brightness-mode (B-mode) images across twelve organs~\cite{jiao2024usfm} and EchoCLIP on over one million echocardiogram images with paired reports~\cite{christensen2024echoclip}, but in both cases, the full training dataset is not publicly available. USF-MAE, on the other hand, was trained on approximately 370k B-mode images pooled from 46 public collections~\cite{megahed2025usfmae}. For echocardiography, there are large datasets of B-mode sequences with corresponding models for functional parameters such as ejection fraction and left ventricular internal diameter~\cite{ouyang2020echonet, echonetlvh}, but these are narrowly scoped to those tasks only. Combining data from several institutions gives more variety in scanners, protocols, and patients than any one group could collect alone, and under a permissive license, both academic and industry researchers can benefit from each other.

Ultrasound is the most widely deployed diagnostic imaging modality in the world~\cite{wangAnatomyFunctionalMolecular2020}. It is real-time, non-ionizing, comparatively inexpensive, and portable enough to be brought to the patient. This matters for the approximately \qty{47}{\percent} of the global population that has little to no access to diagnostics, a gap that is largest at the primary-care level where cost, infrastructure, and staffing constraints rule out \ac{CT} and \ac{MRI}~\cite{fleming2021lancet}. Ultrasound is also the only clinical modality that images at frame rates high enough to resolve millisecond-scale physiology, which underpins shear-wave elastography, functional ultrasound, and super-resolution imaging~\cite{tanter2014ultrafast,errico2015ulm}. However, ultrasound image quality depends on \emph{operator skill} and the \emph{patient's anatomy}, and ultrasound does not offer standard and reproducible imaging planes, which challenges consistent interpretation.

\begin{figure}
    \centering
    \includegraphics[width=0.9\linewidth]{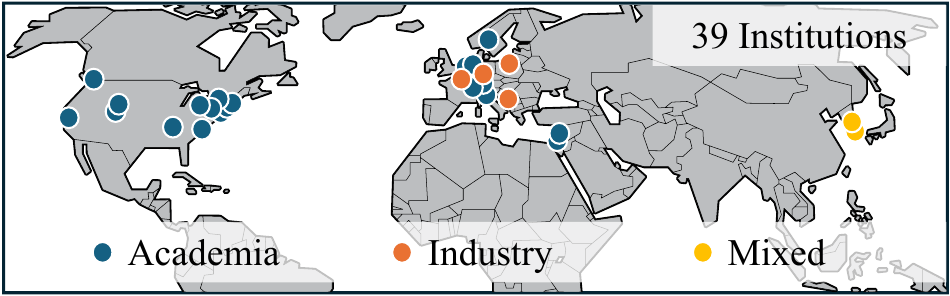}
    \caption{Global map of the origin of the contributions to \OpenHRF, also showing the type of contributor (academia, industry or mixed).}
    \label{fig:world_map}
    \vspace{-16pt}
\end{figure}

Every one of the aforementioned ultrasound datasets consists of brightness-mode (B-mode) images. B-mode is the final product of a lossy pipeline: beamforming, typically under an assumed constant \ac{SoS}, envelope detection, log compression, and vendor-specific post-processing. The raw measurements consist of the \textit{raw channel data}: the echo signals recorded by each individual transducer element after every transmit event (Section~\ref{sec:primer}). Raw channel data retains the channel-level amplitude and phase enabling the analysis of inter-channel coherence, and frequency information that the displayed image discards. Moreover, the dynamic range is typically two orders of magnitude larger than that of B-mode images. Interpretive tasks, such as segmentation, quantification, and diagnosis, can be performed on B-mode images. However, many of the key technical challenges in ultrasound imaging, such as correcting inaccurate assumptions (e.g., aberration correction and reverberation reduction) and tissue characterization (e.g., tomographic attenuation imaging), require access to the raw channel data (Section~\ref{sec:tasks}).

Image reconstruction reduces the data by roughly two orders of magnitude. Beamforming collapses the signals of all elements into a single value per pixel, discarding the per-element information needed to estimate the true \ac{SoS}, correct aberrations, or separate reverberation from the direct echoes, and envelope detection further discards the phase on which Doppler and coherence-based methods rely. Accordingly, a substantial body of work has shown that operating directly on raw channel data or pre-envelope beamformed signals enables capabilities beyond what the displayed image can support~\cite{vansloun2020deep}: adaptive beamforming and image reconstruction~\cite{luijten2020adaptive}, speed-of-sound inversion~\cite{feigin2020soundspeed}, aberration correction~\cite{simsonDifferentiableBeamformingUltrasound2023, simsonUltrasoundAutofocusingCommon2026}, dehazing~\cite{stevensDehazingUltrasoundUsing2024}, microbubble localization for super-resolution imaging~\cite{heiles2022pala}, and clutter filtering and flow estimation~\cite{demene2015clutter, solomon2020deep}. To date, the community has not explored how data scale and variety can support solving these fundamental problems in medical ultrasound imaging.

\begin{figure*}[t]
    \centering
    \includegraphics{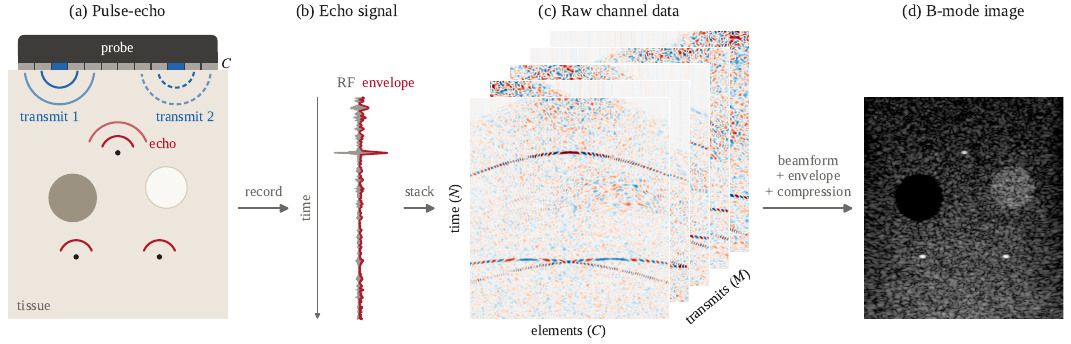}
    \caption{Simplified overview of ultrasound imaging. (a) In each transmit event, the transducer array emits an acoustic pulse into the tissue; shown here are two synthetic aperture transmits, each fired by a single element. Wherever the acoustic properties in the tissue change, part of the wave is scattered back as echoes. (b) Each element records the returning echoes over time as a radio-frequency (\ac{RF}) signal, whose envelope shows the echo strength. (c) Recording all $C$ elements for each of the $M$ transmit events gives one frame of raw channel data with $N$ samples per element ($C\times M\times N$, shown demodulated). Each point scatterer appears as a hyperbola, because its echo reaches each element at a slightly different time. (d) Beamforming aligns and sums these signals for every pixel, and envelope detection and log compression then give the B-mode image. Here we show a simulated phantom with three point targets, an anechoic cyst, and a hyperechoic inclusion.}
    \label{fig:primer}
    \vspace{-8pt}
\end{figure*}

The scarcity of open-access raw channel data can be explained by three factors: \emph{hardware access}, \emph{fragmented formats}, and \emph{data rates}. First, \textit{hardware access}: clinical ultrasound scanners generally do not permit the extraction of raw channel data. Their processing is performed in dedicated hardware created for clinical real-time imaging. Access to raw channel data is commonly considered only useful for device engineering and not readily available.  Furthermore, the raw data rates exceed the design specifications of clinical storage and networking. Raw channel data is therefore available mainly from research scanners requiring dedicated hardware. Second, \textit{fragmented formats}: acquisition parameters and storage conventions vary by vendor and by lab, for example, between the Verasonics MATLAB environment, the us4us ARRUS package, and open initiatives such as the \ac{UFF}~\cite{bernard2018uff}, the \ac{USTB}~\cite{rodriguezmolares2017ustb}, and \zea~\cite{zea2025}. This makes it challenging to combine and explore data from different sources. Third, \textit{data rates}: ultrasound can sample thousands of elements at multi-megahertz rates, producing on the order of a gigabyte per second and terabytes per session~\cite{guan2026mach}. Transferring and storing raw channel data is therefore traditionally burdensome and has imposed a practical limit on the size of  both private and open datasets.

Despite these barriers, several open resources of raw channel data have been released. PICMUS established a standardized benchmark for plane-wave beamforming, consisting of a total of six simulated, phantom, and \emph{in-vivo} acquisitions with common image-quality metrics against which beamformers could be compared~\cite{liebgott2016picmus}. CUBDL pooled and released data from seven institutions and 576 acquisition sequences across \emph{in-vivo}, phantom, and simulated sources, initially intended for deep learning tasks~\cite{hyun2021cubdl,bell2020cubdl,cubdl_data}. PALA and ULMShare made data available for ultrasound localization microscopy, enabling direct comparison of super-resolution and microvascular imaging methods~\cite{heiles2022pala,rauby2026ulmshare}. These efforts demonstrated the value of shared raw channel data. However, each dataset targets a single primary task, covers a limited range of anatomies, probes, and transmit schemes, or uses its own data format. None were designed for, or approach, the scale and variety required for task-agnostic end-to-end foundation model training.

Addressing these shortcomings requires a pooled global initiative as has been demonstrated in other fields. Open X-Embodiment demonstrated that aggregating heterogeneous robot data under a common schema yields models that outperform any of the contributing single-source policies~\cite{openx2023}. Open-H-Embodiment adapted that model to the sensing, safety, and governance constraints of the medical domain, pooling 780 hours of data from more than 50 institutions under a unified format and showing that domain-specific pretraining at that scale transfers across tasks and hardware platforms~\cite{openh2026}. \OpenHRF follows this model for raw channel data. By providing access to a large corpus of raw channel data covering a variety of scanning regimes and imaging media, \OpenHRF aims to support open comparison, algorithmic development, and scientific discovery on raw channel data.

This work introduces \OpenHRF, a community dataset initiative comprising \OpenHRFSize of raw channel data from \OpenHRFAcquisitions acquisitions, spanning \emph{in-vivo}, phantom, and simulated sources (Fig.~\ref{fig:collage}). In terms of \textbf{scale and diversity}, it exceeds the largest pre-existing open collection~\cite{hyun2021cubdl} by more than an order of magnitude and covers a wide range of anatomies, phantoms, probe types, and transmit schemes (Section~\ref{sec:composition}). A \textbf{unified format}, built on existing open standards and the \zea~\cite{zea2025} tooling, makes acquisitions from different scanners, probes, and transmit sequences directly interoperable and documents the conventions needed to interpret them correctly (Section~\ref{sec:data_organization}). \textbf{Permissive licensing} under CC BY 4.0 removes the access agreements and non-commercial restrictions that limit the reuse of several existing ultrasound datasets, allowing both academic and commercial use. Finally, the contributing groups supplied intended \textbf{tasks} with demonstrations, which we organize into five categories: \textit{probe motion}, \textit{tissue motion}, \textit{inaccurate assumptions}, \textit{tissue characterization}, and \textit{therapy} (Section~\ref{sec:tasks}). These include tracked 3D and \ac{SSA} imaging, blood flow imaging, \ac{SoS} estimation, and aberration correction, supported by complementary metadata such as reference probe positions, flow maps, and \ac{SoS} maps stored alongside the raw channel data.

\section{Ultrasound Imaging}
\label{sec:primer}

Ultrasound imaging is based on the pulse-echo principle~\cite{jensen2007medical, vansloun2020deep, perrotSoYouThink2021}, see Fig.~\ref{fig:primer}. A transducer array of $C$ elements emits a short acoustic pulse into the body. We refer to this as a \textit{transmit event}. By firing the elements with relative delays, a wavefront can be focused at a point, steered as a plane wave, diverged from a virtual source behind the array, or emitted from a single element. As the wave propagates, it is partially reflected and scattered wherever the acoustic properties of the tissue change, and the reflected pressure waves return to the transducer as \textit{echoes}. Each element measures the incident pressure integrated over its surface, and this signal is digitized by an analog-to-digital converter at tens of \si{\mega\hertz}, yielding $N$ fast-time samples per element. These digitized pressure measurements are the \textit{raw channel data}, stored either as real-valued \ac{RF} signals or as demodulated, complex-valued \ac{IQ} signals. A \emph{frame} is the set of $M$ transmit events that together are used to reconstruct a single image, so that one frame of raw channel data can be written as a tensor with dimensions $C\times M\times N$. We use the term \textit{acquisition} to refer to a sequence of frames acquired in a single imaging session. Ultrasound imaging is typically two-dimensional, but volumetric (3D) imaging can be achieved using 2D matrix arrays, which enable electronic beam steering and focusing in both the lateral and elevational dimensions \cite{yu2019design}. However, fully sampled matrix arrays require substantially more transducer elements and electronic channels than conventional 1D arrays, resulting in increased hardware complexity and data throughput~\cite{giangrossi2022requirements}.

To form an image, the raw channel data is \textit{beamformed}. The most common method, \ac{DAS}, computes for each pixel the time of flight of the wavefront from the transducer to that pixel and back to each element, and sums the correspondingly delayed channel signals. These delays assume a constant \ac{SoS}, usually \qty{1540}{\meter\per\second}, although soft tissue ranges from about \qty{1450}{\meter\per\second} in fat to above \qty{1600}{\meter\per\second} in muscle. The beamformed signal is then envelope-detected and log-compressed to form the familiar \textit{B-mode} (brightness mode) image. Motion can be measured across multiple frames, which is used in Doppler and vector flow imaging. These imaging modes estimate blood velocity from the phase changes between consecutive frames, after clutter filtering separates the weakly scattering blood from the much stronger tissue signal.

Each image formation step discards information or relies on assumptions that are violated in practice: tissue is heterogeneous in \ac{SoS}, multiple scattering between tissue layers causes reverberation, and attenuation depends on frequency, tissue absorption and scattering. Image formation from raw channel data is therefore an inverse problem with a physically grounded but only approximate forward model. The data is also high-dimensional and high-rate, and the acquisition scheme itself, i.e., which transmit events to fire, is a design choice that can be optimized~\cite{nolanTaskBasedAdaptiveTransmit2026}.

\begin{table}[!tp]
\centering
\begingroup
\fontsize{7}{8.2}\selectfont
\setlength{\tabcolsep}{2pt}
\renewcommand{\arraystretch}{1.04}
\definecolor{openhmediuminvivo}{HTML}{2a78d6}
\definecolor{openhmediumphantom}{HTML}{eb6834}
\definecolor{openhmediumsimulation}{HTML}{1baf7a}
\caption{\OpenHRF contributions: acquisitions by medium and intended tasks.}
\label{tab:dataset-tasks}
\begin{tabularx}{\columnwidth}{@{}l l >{\raggedright\arraybackslash}X@{}}
\toprule
Contribution & Acq. & Tasks \\
\midrule
\texttt{colorado-boulder} & \mbox{\tikz[baseline=-0.75ex]{\fill[openhmediuminvivo] (0.00ex,0.72ex) -- (-0.19ex,0.19ex) -- (-0.76ex,0.17ex) -- (-0.31ex,-0.18ex) -- (-0.47ex,-0.72ex) -- (0.00ex,-0.41ex) -- (0.47ex,-0.72ex) -- (0.31ex,-0.18ex) -- (0.76ex,0.17ex) -- (0.19ex,0.19ex) -- cycle;}\,42}\enspace \mbox{\tikz[baseline=-0.75ex]{\fill[openhmediumphantom] (0,0) circle (0.6ex);}\,20} & Recon, Motion, IQA, 3D \\[1pt]
\texttt{concordia} \cite{sharifzadeh2024} & \mbox{\tikz[baseline=-0.75ex]{\fill[openhmediumsimulation] (-0.68ex,-0.6ex) -- (0.68ex,-0.6ex) -- (0,0.6ex) -- cycle;}\,2,000} & Recon, Interp, QUS \\[1pt]
\texttt{dartmouth-uct} {\fontsize{5.5}{6}\selectfont\cite{badano2018,treeby2010}} & \mbox{\tikz[baseline=-0.75ex]{\fill[openhmediumsimulation] (-0.68ex,-0.6ex) -- (0.68ex,-0.6ex) -- (0,0.6ex) -- cycle;}\,2,625} & Recon, SoS, QUS \\[1pt]
\texttt{jhu-pulselab} {\fontsize{5.5}{6}\selectfont\cite{rodriguezmolares2017ustb}} & \mbox{\tikz[baseline=-0.75ex]{\fill[openhmediumphantom] (0,0) circle (0.6ex);}\,4} & Recon, IQA \\[1pt]
\texttt{kaist-snubh-barreleye} & \mbox{\tikz[baseline=-0.75ex]{\fill[openhmediuminvivo] (0.00ex,0.72ex) -- (-0.19ex,0.19ex) -- (-0.76ex,0.17ex) -- (-0.31ex,-0.18ex) -- (-0.47ex,-0.72ex) -- (0.00ex,-0.41ex) -- (0.47ex,-0.72ex) -- (0.31ex,-0.18ex) -- (0.76ex,0.17ex) -- (0.19ex,0.19ex) -- cycle;}\,70} & Recon, SoS, Interp, QUS \\[1pt]
\texttt{mosaic-nuevosono} & \mbox{\tikz[baseline=-0.75ex]{\fill[openhmediuminvivo] (0.00ex,0.72ex) -- (-0.19ex,0.19ex) -- (-0.76ex,0.17ex) -- (-0.31ex,-0.18ex) -- (-0.47ex,-0.72ex) -- (0.00ex,-0.41ex) -- (0.47ex,-0.72ex) -- (0.31ex,-0.18ex) -- (0.76ex,0.17ex) -- (0.19ex,0.19ex) -- cycle;}\,9} & Interp, Motion \\[1pt]
\texttt{nv-raw2insights-us} {\fontsize{5.5}{6}\selectfont\cite{nv_raw2insights_us}} & \mbox{\tikz[baseline=-0.75ex]{\fill[openhmediumsimulation] (-0.68ex,-0.6ex) -- (0.68ex,-0.6ex) -- (0,0.6ex) -- cycle;}\,923} & Recon, SoS, Interp \\[1pt]
\texttt{oslo} {\fontsize{5.5}{6}\selectfont\cite{rodriguezmolares2017ustb}} & \mbox{\tikz[baseline=-0.75ex]{\fill[openhmediuminvivo] (0.00ex,0.72ex) -- (-0.19ex,0.19ex) -- (-0.76ex,0.17ex) -- (-0.31ex,-0.18ex) -- (-0.47ex,-0.72ex) -- (0.00ex,-0.41ex) -- (0.47ex,-0.72ex) -- (0.31ex,-0.18ex) -- (0.76ex,0.17ex) -- (0.19ex,0.19ex) -- cycle;}\,5}\enspace \mbox{\tikz[baseline=-0.75ex]{\fill[openhmediumphantom] (0,0) circle (0.6ex);}\,19}\enspace \mbox{\tikz[baseline=-0.75ex]{\fill[openhmediumsimulation] (-0.68ex,-0.6ex) -- (0.68ex,-0.6ex) -- (0,0.6ex) -- cycle;}\,11} & Recon, Motion, IQA \\[1pt]
\texttt{politorino} {\fontsize{5.5}{6}\selectfont\cite{politorino}} & \mbox{\tikz[baseline=-0.75ex]{\fill[openhmediuminvivo] (0.00ex,0.72ex) -- (-0.19ex,0.19ex) -- (-0.76ex,0.17ex) -- (-0.31ex,-0.18ex) -- (-0.47ex,-0.72ex) -- (0.00ex,-0.41ex) -- (0.47ex,-0.72ex) -- (0.31ex,-0.18ex) -- (0.76ex,0.17ex) -- (0.19ex,0.19ex) -- cycle;}\,30} & Motion \\[1pt]
\texttt{resolvestroke} & \mbox{\tikz[baseline=-0.75ex]{\fill[openhmediuminvivo] (0.00ex,0.72ex) -- (-0.19ex,0.19ex) -- (-0.76ex,0.17ex) -- (-0.31ex,-0.18ex) -- (-0.47ex,-0.72ex) -- (0.00ex,-0.41ex) -- (0.47ex,-0.72ex) -- (0.31ex,-0.18ex) -- (0.76ex,0.17ex) -- (0.19ex,0.19ex) -- cycle;}\,40}\enspace \mbox{\tikz[baseline=-0.75ex]{\fill[openhmediumphantom] (0,0) circle (0.6ex);}\,3} & Recon, Motion, IQA, 3D, Flow \\[1pt]
\texttt{siemens-healthineers} & \mbox{\tikz[baseline=-0.75ex]{\fill[openhmediuminvivo] (0.00ex,0.72ex) -- (-0.19ex,0.19ex) -- (-0.76ex,0.17ex) -- (-0.31ex,-0.18ex) -- (-0.47ex,-0.72ex) -- (0.00ex,-0.41ex) -- (0.47ex,-0.72ex) -- (0.31ex,-0.18ex) -- (0.76ex,0.17ex) -- (0.19ex,0.19ex) -- cycle;}\,590} & Recon \\[1pt]
\texttt{stanford-murine} {\fontsize{5.5}{6}\selectfont\cite{stanford_murine_data,stanford_phantom_data,telichko2022,ali2022}} & \mbox{\tikz[baseline=-0.75ex]{\fill[openhmediuminvivo] (0.00ex,0.72ex) -- (-0.19ex,0.19ex) -- (-0.76ex,0.17ex) -- (-0.31ex,-0.18ex) -- (-0.47ex,-0.72ex) -- (0.00ex,-0.41ex) -- (0.47ex,-0.72ex) -- (0.31ex,-0.18ex) -- (0.76ex,0.17ex) -- (0.19ex,0.19ex) -- cycle;}\,102}\enspace \mbox{\tikz[baseline=-0.75ex]{\fill[openhmediumphantom] (0,0) circle (0.6ex);}\,143} & Recon, SoS \\[1pt]
\texttt{strasbourg-basel} & \mbox{\tikz[baseline=-0.75ex]{\fill[openhmediumphantom] (0,0) circle (0.6ex);}\,9} & Recon, SoS, 3D \\[1pt]
\texttt{technion} & \mbox{\tikz[baseline=-0.75ex]{\fill[openhmediuminvivo] (0.00ex,0.72ex) -- (-0.19ex,0.19ex) -- (-0.76ex,0.17ex) -- (-0.31ex,-0.18ex) -- (-0.47ex,-0.72ex) -- (0.00ex,-0.41ex) -- (0.47ex,-0.72ex) -- (0.31ex,-0.18ex) -- (0.76ex,0.17ex) -- (0.19ex,0.19ex) -- cycle;}\,39}\enspace \mbox{\tikz[baseline=-0.75ex]{\fill[openhmediumphantom] (0,0) circle (0.6ex);}\,1} & Recon, Motion, IQA, Clutter \\[1pt]
\texttt{tel-aviv} {\fontsize{5.5}{6}\selectfont\cite{grutman2025,ma2024}} & \mbox{\tikz[baseline=-0.75ex]{\fill[openhmediuminvivo] (0.00ex,0.72ex) -- (-0.19ex,0.19ex) -- (-0.76ex,0.17ex) -- (-0.31ex,-0.18ex) -- (-0.47ex,-0.72ex) -- (0.00ex,-0.41ex) -- (0.47ex,-0.72ex) -- (0.31ex,-0.18ex) -- (0.76ex,0.17ex) -- (0.19ex,0.19ex) -- cycle;}\,16}\enspace \mbox{\tikz[baseline=-0.75ex]{\fill[openhmediumphantom] (0,0) circle (0.6ex);}\,6} & Interp \\[1pt]
\texttt{tue-aaa} & \mbox{\tikz[baseline=-0.75ex]{\fill[openhmediuminvivo] (0.00ex,0.72ex) -- (-0.19ex,0.19ex) -- (-0.76ex,0.17ex) -- (-0.31ex,-0.18ex) -- (-0.47ex,-0.72ex) -- (0.00ex,-0.41ex) -- (0.47ex,-0.72ex) -- (0.31ex,-0.18ex) -- (0.76ex,0.17ex) -- (0.19ex,0.19ex) -- cycle;}\,15} & Recon, Motion \\[1pt]
\texttt{tue-cardiac} {\fontsize{5.5}{6}\selectfont\cite{penninga2026refocus}} & \mbox{\tikz[baseline=-0.75ex]{\fill[openhmediuminvivo] (0.00ex,0.72ex) -- (-0.19ex,0.19ex) -- (-0.76ex,0.17ex) -- (-0.31ex,-0.18ex) -- (-0.47ex,-0.72ex) -- (0.00ex,-0.41ex) -- (0.47ex,-0.72ex) -- (0.31ex,-0.18ex) -- (0.76ex,0.17ex) -- (0.19ex,0.19ex) -- cycle;}\,12} & Recon \\[1pt]
\texttt{tue-carotid} & \mbox{\tikz[baseline=-0.75ex]{\fill[openhmediuminvivo] (0.00ex,0.72ex) -- (-0.19ex,0.19ex) -- (-0.76ex,0.17ex) -- (-0.31ex,-0.18ex) -- (-0.47ex,-0.72ex) -- (0.00ex,-0.41ex) -- (0.47ex,-0.72ex) -- (0.31ex,-0.18ex) -- (0.76ex,0.17ex) -- (0.19ex,0.19ex) -- cycle;}\,80} & Recon, IQA \\[1pt]
\texttt{tumunich} & \mbox{\tikz[baseline=-0.75ex]{\fill[openhmediumphantom] (0,0) circle (0.6ex);}\,11} & Recon, 3D \\[1pt]
\texttt{twente-cavitation} & \mbox{\tikz[baseline=-0.75ex]{\fill[openhmediumphantom] (0,0) circle (0.6ex);}\,19} & MCD \\[1pt]
\texttt{twente-microbubblesim} {\fontsize{5.5}{6}\selectfont\cite{twente_microbubblesim}} & \mbox{\tikz[baseline=-0.75ex]{\fill[openhmediumsimulation] (-0.68ex,-0.6ex) -- (0.68ex,-0.6ex) -- (0,0.6ex) -- cycle;}\,500} & Flow \\[1pt]
\texttt{twente-vortexflow} & \mbox{\tikz[baseline=-0.75ex]{\fill[openhmediumphantom] (0,0) circle (0.6ex);}\,6} & Recon, IQA, Flow \\[1pt]
\texttt{ubc} {\fontsize{5.5}{6}\selectfont\cite{ubc_module_c}} & \mbox{\tikz[baseline=-0.75ex]{\fill[openhmediumsimulation] (-0.68ex,-0.6ex) -- (0.68ex,-0.6ex) -- (0,0.6ex) -- cycle;}\,7,845} & Recon, Interp, 3D \\[1pt]
\texttt{ulmshare} {\fontsize{5.5}{6}\selectfont\cite{ulmshare_data,rauby2026ulmshare}} & \mbox{\tikz[baseline=-0.75ex]{\fill[openhmediuminvivo] (0.00ex,0.72ex) -- (-0.19ex,0.19ex) -- (-0.76ex,0.17ex) -- (-0.31ex,-0.18ex) -- (-0.47ex,-0.72ex) -- (0.00ex,-0.41ex) -- (0.47ex,-0.72ex) -- (0.31ex,-0.18ex) -- (0.76ex,0.17ex) -- (0.19ex,0.19ex) -- cycle;}\,99} & Motion, Flow, Clutter \\[1pt]
\texttt{unc-liver} {\fontsize{5.5}{6}\selectfont\cite{zhuang2026,sode2026fullwave,sode2026attenuation}} & \mbox{\tikz[baseline=-0.75ex]{\fill[openhmediumsimulation] (-0.68ex,-0.6ex) -- (0.68ex,-0.6ex) -- (0,0.6ex) -- cycle;}\,1,906} & Recon, SoS, Interp, Clutter \\[1pt]
\texttt{unc-openpros} {\fontsize{5.5}{6}\selectfont\cite{unc_openpros}} & \mbox{\tikz[baseline=-0.75ex]{\fill[openhmediumsimulation] (-0.68ex,-0.6ex) -- (0.68ex,-0.6ex) -- (0,0.6ex) -- cycle;}\,248} & Recon, SoS, QUS \\[1pt]
\texttt{us4us} & \mbox{\tikz[baseline=-0.75ex]{\fill[openhmediuminvivo] (0.00ex,0.72ex) -- (-0.19ex,0.19ex) -- (-0.76ex,0.17ex) -- (-0.31ex,-0.18ex) -- (-0.47ex,-0.72ex) -- (0.00ex,-0.41ex) -- (0.47ex,-0.72ex) -- (0.31ex,-0.18ex) -- (0.76ex,0.17ex) -- (0.19ex,0.19ex) -- cycle;}\,16}\enspace \mbox{\tikz[baseline=-0.75ex]{\fill[openhmediumphantom] (0,0) circle (0.6ex);}\,2} & Recon \\[1pt]
\texttt{vanderbilt} & \mbox{\tikz[baseline=-0.75ex]{\fill[openhmediuminvivo] (0.00ex,0.72ex) -- (-0.19ex,0.19ex) -- (-0.76ex,0.17ex) -- (-0.31ex,-0.18ex) -- (-0.47ex,-0.72ex) -- (0.00ex,-0.41ex) -- (0.47ex,-0.72ex) -- (0.31ex,-0.18ex) -- (0.76ex,0.17ex) -- (0.19ex,0.19ex) -- cycle;}\,164}\enspace \mbox{\tikz[baseline=-0.75ex]{\fill[openhmediumphantom] (0,0) circle (0.6ex);}\,1} & Recon, IQA, Clutter \\[1pt]
\texttt{waterloo-carotid} {\fontsize{5.5}{6}\selectfont\cite{waterloo_carotid}} & \mbox{\tikz[baseline=-0.75ex]{\fill[openhmediuminvivo] (0.00ex,0.72ex) -- (-0.19ex,0.19ex) -- (-0.76ex,0.17ex) -- (-0.31ex,-0.18ex) -- (-0.47ex,-0.72ex) -- (0.00ex,-0.41ex) -- (0.47ex,-0.72ex) -- (0.31ex,-0.18ex) -- (0.76ex,0.17ex) -- (0.19ex,0.19ex) -- cycle;}\,93} & Recon, Motion, Flow, Clutter \\[1pt]
\texttt{waterloo-femoralvein} {\fontsize{5.5}{6}\selectfont\cite{waterloo_femoralvein}} & \mbox{\tikz[baseline=-0.75ex]{\fill[openhmediuminvivo] (0.00ex,0.72ex) -- (-0.19ex,0.19ex) -- (-0.76ex,0.17ex) -- (-0.31ex,-0.18ex) -- (-0.47ex,-0.72ex) -- (0.00ex,-0.41ex) -- (0.47ex,-0.72ex) -- (0.31ex,-0.18ex) -- (0.76ex,0.17ex) -- (0.19ex,0.19ex) -- cycle;}\,15} & Recon, Motion, Flow, Clutter \\[1pt]
\texttt{waterloo-muscle} {\fontsize{5.5}{6}\selectfont\cite{waterloo_muscle}} & \mbox{\tikz[baseline=-0.75ex]{\fill[openhmediuminvivo] (0.00ex,0.72ex) -- (-0.19ex,0.19ex) -- (-0.76ex,0.17ex) -- (-0.31ex,-0.18ex) -- (-0.47ex,-0.72ex) -- (0.00ex,-0.41ex) -- (0.47ex,-0.72ex) -- (0.31ex,-0.18ex) -- (0.76ex,0.17ex) -- (0.19ex,0.19ex) -- cycle;}\,1,248} & Recon, SoS \\[1pt]
\texttt{weillcornell} & \mbox{\tikz[baseline=-0.75ex]{\fill[openhmediumphantom] (0,0) circle (0.6ex);}\,120} & Recon, QUS \\[1pt]
\texttt{weizmann-sampl} & \mbox{\tikz[baseline=-0.75ex]{\fill[openhmediuminvivo] (0.00ex,0.72ex) -- (-0.19ex,0.19ex) -- (-0.76ex,0.17ex) -- (-0.31ex,-0.18ex) -- (-0.47ex,-0.72ex) -- (0.00ex,-0.41ex) -- (0.47ex,-0.72ex) -- (0.31ex,-0.18ex) -- (0.76ex,0.17ex) -- (0.19ex,0.19ex) -- cycle;}\,30} & Recon, SoS \\[1pt]
\texttt{wpi} {\fontsize{5.5}{6}\selectfont\cite{murakami2025}} & \mbox{\tikz[baseline=-0.75ex]{\fill[openhmediumphantom] (0,0) circle (0.6ex);}\,5}\enspace \mbox{\tikz[baseline=-0.75ex]{\fill[openhmediumsimulation] (-0.68ex,-0.6ex) -- (0.68ex,-0.6ex) -- (0,0.6ex) -- cycle;}\,190} & Recon, 3D \\[1pt]
\midrule
\textbf{Total (\OpenHRFContributors contributions)} & \rlap{\mbox{\tikz[baseline=-0.75ex]{\fill[openhmediuminvivo] (0.00ex,0.72ex) -- (-0.19ex,0.19ex) -- (-0.76ex,0.17ex) -- (-0.31ex,-0.18ex) -- (-0.47ex,-0.72ex) -- (0.00ex,-0.41ex) -- (0.47ex,-0.72ex) -- (0.31ex,-0.18ex) -- (0.76ex,0.17ex) -- (0.19ex,0.19ex) -- cycle;}\,2,715}\enspace \mbox{\tikz[baseline=-0.75ex]{\fill[openhmediumphantom] (0,0) circle (0.6ex);}\,369}\enspace \mbox{\tikz[baseline=-0.75ex]{\fill[openhmediumsimulation] (-0.68ex,-0.6ex) -- (0.68ex,-0.6ex) -- (0,0.6ex) -- cycle;}\,16,248}} & \hfill \OpenHRFAcquisitions acquisitions \\
\bottomrule
\end{tabularx}
\par\smallskip
\begin{minipage}{\columnwidth}\fontsize{7}{8.2}\selectfont\raggedright
\textbf{Media:} \tikz[baseline=-0.75ex]{\fill[openhmediuminvivo] (0.00ex,0.72ex) -- (-0.19ex,0.19ex) -- (-0.76ex,0.17ex) -- (-0.31ex,-0.18ex) -- (-0.47ex,-0.72ex) -- (0.00ex,-0.41ex) -- (0.47ex,-0.72ex) -- (0.31ex,-0.18ex) -- (0.76ex,0.17ex) -- (0.19ex,0.19ex) -- cycle;} \emph{in-vivo}; \ \tikz[baseline=-0.75ex]{\fill[openhmediumphantom] (0,0) circle (0.6ex);} phantom; \ \tikz[baseline=-0.75ex]{\fill[openhmediumsimulation] (-0.68ex,-0.6ex) -- (0.68ex,-0.6ex) -- (0,0.6ex) -- cycle;} simulation. \par
\textbf{Tasks:} \textbf{Recon}: reconstruction, beamforming; \textbf{SoS}: Speed of sound estimation, aberration correction; \textbf{Interp}: interpretation, segmentation, detection; \textbf{Motion}: Motion tracking; \textbf{IQA}: Image quality assessment; \textbf{3D}: tracked reconstruction, matrix imaging; \textbf{Flow}: Blood flow, ULM; \textbf{QUS}: Quantitative ultrasound; \textbf{Clutter}: Clutter/reverb suppression; \textbf{MCD}: Microbubble cavitation detection.
\end{minipage}
\endgroup
\vspace{-16pt}
\end{table}

\section{Methods}
\label{sec:methods}
\subsection{Call for Contributions}
\label{sec:collection}
Data collection began with a public announcement and \ac{RFP} document. The \ac{RFP} described the motivation for collecting a dataset of raw channel data and subsequent training of a foundation model, along with a list of task groups for which access to raw channel data is particularly relevant. The \ac{RFP} specified that the provided data must be channel data, and provided a data tier list ranging from \emph{in-vivo} human data to digitally simulated data. An important requirement for each contribution was that its data fell under the Creative Commons BY 4.0 license (CC BY 4.0), ensuring that it could be freely shared and reused provided appropriate attribution was given. For \emph{in-vivo} data, the ethical approval under which it was acquired is documented in the data card of each contribution, and all protected health information was removed so that the data are de-identified. Accepted proposals were then provided with a submission guide, which instructed that the data must be supplied in \zea format (Section~\ref{sec:data_organization}), along with a basic reconstruction script to visualize the data, and a license file.
After each initial contribution, the \OpenHRF core team reviewed the submitted dataset to spot any spurious or missing metadata entries and to check whether the reconstruction script would indeed produce the expected result. Contributors were also required to provide a reference image against which the data curators could validate reproducibility. Data were collected globally from \OpenHRFInstitutions institutions (Fig.~\ref{fig:world_map}). Contributors across academia and industry created a total of \OpenHRFContributors contributions (Table \ref{tab:dataset-tasks}).

\subsection{Task Categories}
\label{sec:tasks}
\noindent Contributions can be sorted into five general categories: probe motion compensation, tissue motion estimation, correction of inaccurate image formation assumptions, tissue characterization, and therapy. In the following, we describe each task in the context of these categories, explaining how access to channel data supports research on that task.

\subsubsection{Probe motion}
Conventional reconstruction treats each frame as an independent 2D slice acquired from a fixed probe position. When the probe is moved, knowledge of its pose allows measurements from different positions to be combined. \textbf{3D / tracked reconstruction (3D):} Reconstructing volumes from tracked or mechanically moved probes, or from matrix arrays. This allows organs, tumors, and fractures to be measured in 3D. Similarly, when the probe moves in-plane, every frame observes the same scatterers from a new position, and the frames can be summed in phase to synthesize an aperture that is larger than the array by itself. Phase-preserving measurements permit coherent synthetic-aperture focusing across probe positions.

\begin{figure*}[!t]
    \centering
    \begin{minipage}[t]{\columnwidth}
        \centering
        \includegraphics[width=0.95\linewidth]{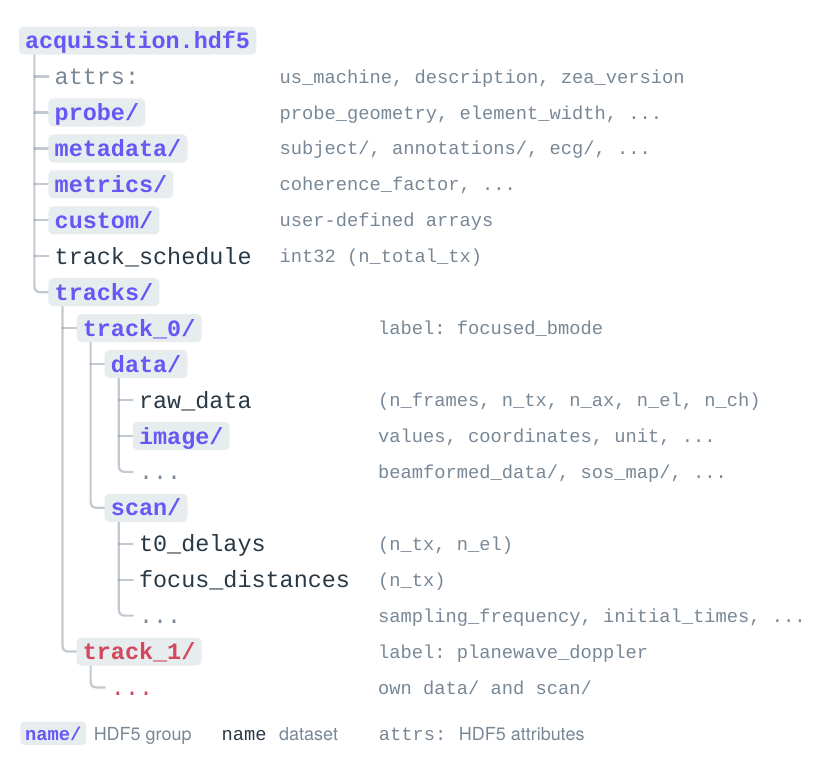}
        \caption{Layout of a \zea acquisition file. Channel data and parameters required to reconstruct it are kept together in a single HDF5 file. Interleaved transmit sequences are stored as separate \emph{tracks} that share the probe and metadata of the file; single-track files place \texttt{data/} and \texttt{scan/} directly at the root. Ellipses denote further fields of the specification, which are documented in full in the \zea docs.}
        \label{fig:spec}
    \end{minipage}\hfill
    \begin{minipage}[t]{\columnwidth}
        \centering
        \includegraphics[width=0.8\linewidth]{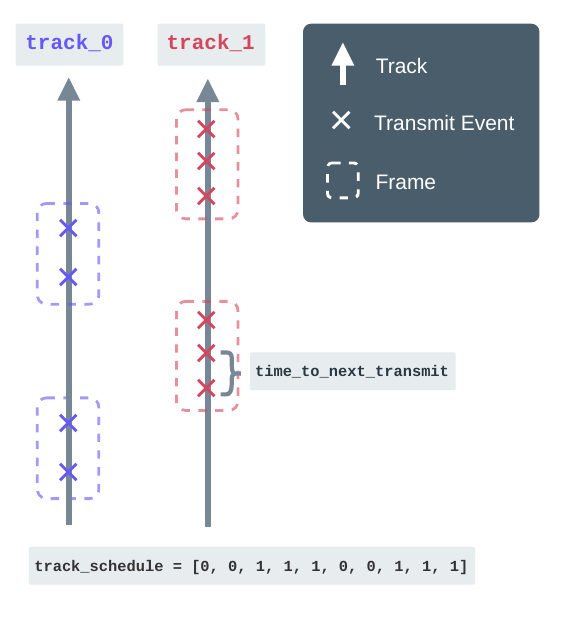}
        \caption{Illustrative example of a \zea file with two tracks with interleaving transmits. A global track schedule records which track each transmit belongs to, so the full acquisition timing can be reconstructed across tracks. Tracks allow different types of transmit events to be stored in the same acquisition file. Generally, each track has its own \texttt{scan} parameters and is processed by its own pipeline.}
        \label{fig:tracks}
    \end{minipage}
\end{figure*}

\subsubsection{Tissue motion}
Moving tissue and flowing blood change the received signals from one transmit to the next. This motion can be a source of diagnostic information, but can also produce noise that must be filtered during reconstruction. \textbf{Motion estimation / tracking (Motion):} Estimating tissue displacement~\cite{ashikuzzaman2024}, for example via shear wave elastography, following structures such as muscle fascicles over time, or aneurysm wall mechanics as an indicator of rupture risk. Access to channel data adds control over focusing, enables phase-sensitive estimation of lower-velocity flows, and allows motion estimation and reconstruction to be performed jointly. \textbf{Blood flow / ultrasound localization microscopy (Flow):} Estimating blood motion with Doppler methods, or localizing and tracking microbubbles to map the microvasculature. Applications include transcranial imaging in stroke, carotid and venous hemodynamics~\cite{waterloo_carotid,waterloo_femoralvein}, and micrometer-scale cerebral microvasculature~\cite{ulmshare_data}. Channel data access enables optimizing beamforming, angle compounding, and tissue suppression, in addition to estimation or localization.

\subsubsection{Inaccurate assumptions}
Delay-and-sum beamforming relies on simplified propagation assumptions, including a known homogeneous background sound speed and the Born approximation, thereby neglecting effects such as refraction and multiple scattering. These inaccuracies result in image quality degradation, motivating the development of methods with better assumptions. \textbf{Generalized reconstruction / beamforming (Recon):} Image formation from the measured signals using delay-and-sum, adaptive, model-based, or learned methods, including tomographic reconstruction from ring and limited-view arrays. Access to channel data before summation allows different reconstruction methods to be compared on the same acquisition. Additionally, it allows computing channel-domain metrics such as \ac{LOC} for more elaborate evaluation~\cite{longLagOneCoherence2018}.  \textbf{Aberration correction (SoS):} Correcting the defocusing caused by an incorrect assumed sound speed. Aberration from fat and muscle layers is a main cause of poor image quality in difficult-to-image patients~\cite{ali2023aberration}. Channel data preserve the arrival times and phase across the aperture needed to estimate and compensate these propagation errors. \textbf{Clutter / reverberation suppression (Clutter):} The removal of unwanted contributions such as stationary tissue, off-axis echoes, or multi-path echoes is a major area of research in ultrasound imaging. Clutter suppression is particularly relevant to cardiac and abdominal imaging, where body-wall reverberation degrades images~\cite{lediju2008clutter, fatemi2019reverberation, stevensDehazingUltrasoundUsing2024}, and in Doppler and ULM, where tissue motion must be suppressed~\cite{demene2015clutter}. Aperture-domain methods exploit differences in coherence and arrival structure across elements, which are lost after summation. \textbf{Image-quality assessment (IQA):} Quantifying resolution, contrast, and artifact levels, and hence the effect of the corrections above, often requires channel data to evaluate alternative beamformers on identical measurements.

\subsubsection{Tissue characterization}
Beyond depicting anatomy, many tasks aim to image and interpret the physical properties of tissue. \textbf{Speed-of-sound estimation (SoS):} Estimating the speed of sound in the medium. Sound speed decreases with hepatic fat content in fatty liver disease~\cite{telichko2022} and is linked to muscle health~\cite{waterloo_muscle}. Channel data preserve the arrival times and phase across the aperture, which allows computing phase-based measures of signal coherence such as \ac{CMPE}~\cite{simsonUltrasoundAutofocusingCommon2026}, which sound-speed estimation methods rely on. \textbf{Quantitative ultrasound (QUS):} Estimating tissue properties such as attenuation, backscatter coefficients, or echo-envelope statistics, which allow distinguishing benign from malignant breast masses and grading liver steatosis. Channel data provides access to the frequency content of the backscattered signal, which log-compressed B-mode images do not preserve. \textbf{Segmentation / interpretation (Interp):} Identifying and/or delineating anatomical structures, or classifying findings, in measurements or images. Applications include breast lesion diagnosis, tumor volumetry, vessel-wall delineation in intravascular imaging, and standard-view recognition in echocardiography and fetal imaging. While conventional image-based methods do not rely on channel data, these contributions enable the development of direct raw-to-insight models and studies of how reconstruction choices affect downstream performance.

\subsubsection{Therapy}
Beyond imaging, ultrasound is used therapeutically, for example to transiently open the blood-brain barrier for targeted drug delivery~\cite{hynynen2001bbb}. \textbf{Microbubble cavitation detection (MCD):} Detecting and characterizing acoustic emissions from bubble activity during ultrasound exposure. Monitoring this bubble activity underpins the safety and efficacy of ultrasound drug delivery and blood-brain-barrier opening. Recorded waveforms with sufficient bandwidth are required, and array data further enable passive mapping of emission locations.

\begin{figure}
    \centering
    \includegraphics[width=1\linewidth]{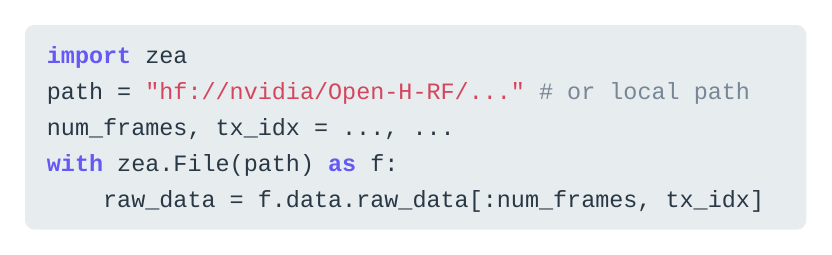}
    \caption{Streaming a portion of a \zea file directly from the \OpenHRF corpus. Accessing only a subset of frames or transmits streams and caches only that part of the data to disk.}
    \label{fig:zea-streaming}
\end{figure}
\subsection{Data Organization}
\label{sec:data_organization}
One of the central tasks of the \OpenHRF initiative is to compile the contributions of all teams into a single, unified format, together with the tooling needed to make the resulting corpus readily accessible, both for a casual inspector and for demanding workflows such as model training. The format builds on the \zea file specification~\cite{zea2025}, which was extended in collaboration with the \OpenHRF initiative and is summarized in Fig.~\ref{fig:spec}.

\subsubsection{File format}
Raw channel data is a digitized record of the pressure field, integrated over the surface of each element (Section~\ref{sec:primer}). Its samples can only be interpreted together with a description of the transducer elements that measured this field and of the transmit events that produced it. Each file therefore holds a single acquisition together with everything needed to process it, organized in a fixed hierarchy of groups (Fig.~\ref{fig:spec}): a \texttt{data} group containing the raw channel data; a \texttt{scan} group containing the acquisition parameters necessary for reconstruction, such as the transmit delays, the sampling frequency, and the start time of each receive window; and a \texttt{probe} group storing the transducer geometry and properties, such as the position and width of every element. Optional groups store provenance and annotations (\texttt{metadata}), quality \texttt{metrics}, and contributor-specific arrays (\texttt{custom}).

The task categories of Section~\ref{sec:tasks} place different demands on the data, yet they share the same underlying measurement: the signal each element recorded after each transmit event. The format therefore separates this measurement from task-specific reference data, which is stored in the same file. The \texttt{data} group holds the raw channel data together with derived spatial maps, each stored with its coordinates and unit, such as beamformed data, attenuation maps, images, or the \ac{SoS} maps used to evaluate aberration correction. Other references, such as probe positions for tracked reconstruction, true microbubble positions for localization microscopy, or segmentations, are stored alongside the raw channel data as well.

Yet, contributions differ not only in their parameters and references, but also in how they organize an acquisition. A single acquisition may interleave several transmit sequences, for example focused transmits for B-mode imaging with plane waves for Doppler. The format represents each sequence in an acquisition as a \emph{track}: a labeled pair of \texttt{data} and \texttt{scan} groups within the same file. The \texttt{probe}, \texttt{metadata}, and \texttt{metrics} groups are shared across tracks, as they belong to the same acquisition. A global track schedule records which track each transmit belongs to, so that the timing of the full acquisition can be reconstructed (Fig.~\ref{fig:tracks}). Each track can thus be processed by its own pipeline, while all the data related to a single acquisition is kept together in one file.

\subsubsection{Data access}
Even with all parameters stored, accessing raw channel data requires \emph{domain knowledge}: reconstructing an image involves a chain of processing steps, such as filtering, demodulation, beamforming, and envelope detection, whose order and settings differ between contributions. We therefore asked each contributor to provide a reproducible reconstruction pipeline together with a reference image, against which we verified that the pipeline produces the intended result (Section~\ref{sec:collection}). This pipeline is stored with the contribution and specifies the exact reconstruction steps applied to its data, and each file records the version of the reconstruction library (\zea~\cite{zea2025}) with which it was stored. Together, these allow any recipient to use the data exactly as intended by the original authors. Moreover, the pipelines are implemented with \ac{GPU}-accelerated, differentiable operations in a machine learning framework, so that ultrasound reconstruction can be integrated directly into model training. The corpus itself is versioned as well: \OpenHRF is hosted at \url{https://huggingface.co/datasets/nvidia/Open-H-RF}, this paper describes \texttt{v1.1}, and any release can be directly loaded through \zea by its tag, so follow-up work can state exactly which version of the corpus was used.

The \emph{data rates} associated with raw channel data pose a second barrier to access (Section~\ref{sec:primer}). An acquisition of a few frames quickly reaches multiple gigabytes, and the entire corpus spans \OpenHRFSize. Downloading the corpus in full is therefore impractical for most users. Instead, users can filter the corpus, for example by anatomy, probe type, or intended task, and access only the parts they need. Within each file, the raw channel data is split into compressed chunks along the transmit axis, so that selected frames or transmits can be read locally or streamed remotely without loading the full acquisition, see Fig.~\ref{fig:zea-streaming} for an example.

\begin{figure}[t]
    \centering
    \includegraphics[width=\linewidth]{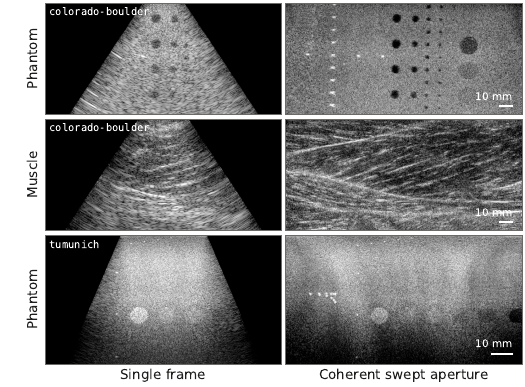}
    \caption{Tracked \acs{SSA}. The \texttt{colorado-boulder} and \texttt{tumunich} contributions demonstrate an \acs{SSA} (right column) built from multiple individual acquisitions (left column).}
    \label{fig:ssa}
\end{figure}

\section{Results}
In the following, we demonstrate several of the tasks introduced above and how they differ across contributions. Then, we provide an analysis of dataset composition, quantifying its overall size, and breakdowns in terms of anatomy, probe type, and transmit scheme.

\subsection{Task Examples}
\label{sec:examples}

\subsubsection{3D / tracked reconstruction (3D)}

Six of the \OpenHRFContributors contributions record where each frame was acquired, alongside the raw channel data.
The position comes from a robot (\texttt{tumunich}), optical tracking (\texttt{colorado-boulder}, NDI Polaris Vega XT), a robot-mounted electromagnetic sensor (\texttt{strasbourg-basel}, trakSTAR), a motorized rotation stage (\texttt{wpi}), a catheter pullback encoder (\texttt{mosaic-nuevosono}), or a positioning system that steps the subject through a ring array (\texttt{us4us}).
Tracking can turn a small acquisition into a larger 2D or 3D one.

\begin{figure}[t]
    \centering
    \includegraphics[width=\linewidth]{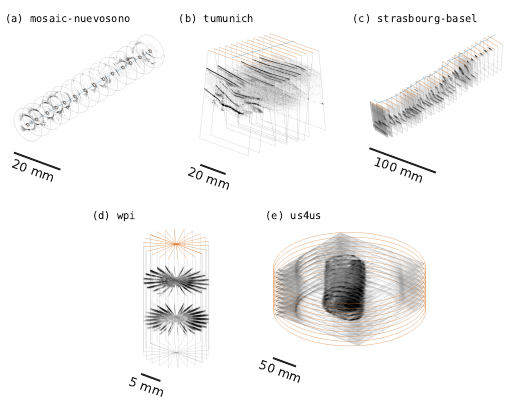}
    \caption{Variety of 3D acquisition geometries in \OpenHRF. Each plane is one acquired frame at its recorded position, with its outline; orange marks the transducer elements (the rotating element at the catheter center in (a), the array face in (b)--(d), the ring in (e)), and the blue line is the tracked probe path.
    (a)~\texttt{mosaic-nuevosono}: \acs{IVUS} catheter pullback in an \emph{in-vivo} porcine artery (14 of 100 cross-sections over \qty{67}{\mm}).
    (b)~\texttt{tumunich}: robot-tracked elevational sweep of a thyroid phantom (12 of \num{1237} frames).
    (c)~\texttt{strasbourg-basel}: robot-tracked sweep along a fractured femur phantom on a curved \qty{276}{\mm} path (32 of 158 frames).
    (d)~\texttt{wpi}: 1D array rotated \ang{181} about its axial axis over a phantom (12 of 182 frames).
    (e)~\texttt{us4us}: ring-array \acs{USCT} of an \emph{in-vivo} forearm, 10 slices \qty{9}{\mm} apart; transmission-attenuation reconstructions rather than B-modes.}
    \label{fig:slices3d}
\end{figure}

\begin{table}[b]
    \centering
    \caption{Lateral width (\qty{-6}{\dB}) of the wire targets under coherent and incoherent compounding.}
    \label{tab:wire_targets_ssa}
    \begin{tabular}{lccc}
    \toprule
    Dataset & Wire targets & \multicolumn{2}{c}{\qty{-6}{\dB} lateral width [\si{\mm}]} \\
    \cmidrule(lr){3-4}
     & & Coherent & Incoherent \\
    \midrule
    \texttt{colorado-boulder} & 12 & \num{0.58 \pm 0.28} & \num{1.96 \pm 0.36} \\
    \texttt{tumunich} & 9 & \num{0.29 \pm 0.05} & \num{0.51 \pm 0.18} \\
    \bottomrule
    \end{tabular}
\end{table}

\paragraph{Swept Synthetic Aperture}

Two contributions contain tracked, in-plane movement, enabling swept synthetic aperture reconstructions (Fig.~\ref{fig:ssa}): \texttt{colorado-boulder} sweeps \qty{160}{\mm} with a \qty{20}{\mm} array, and \texttt{tumunich} \qty{70}{\mm} with a \qty{38}{\mm} array.
When measured on the wire targets of their phantoms, coherent summation narrows the \qty{-6}{\dB} lateral width by \num{3.4}$\times$ and \num{1.8}$\times$, respectively (Table~\ref{tab:wire_targets_ssa}).

\begin{figure}[ht]
    \centering
    \includegraphics[width=\linewidth]{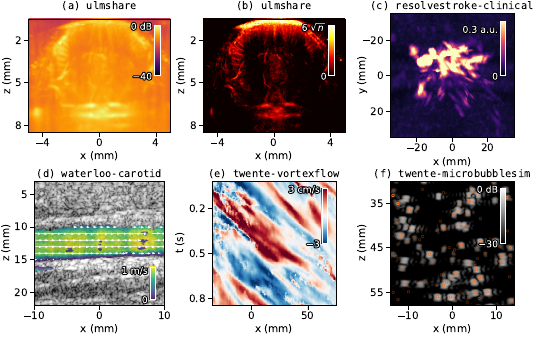}
    \caption{Blood flow in \OpenHRF.
    (a), (b)~\texttt{ulmshare}: mouse brain through the intact skull, from \num{20000} frames: power Doppler after singular-value clutter filtering (a) and \acs{ULM} track density at \qty{10}{\um} (b).
    (c)~\texttt{resolvestroke-clinical}: top-view maximum-intensity projection of the stored 3D microvascular image of a human brain.
    (d)~\texttt{waterloo-carotid}: vector velocity in a human common carotid artery at peak systole, drawn where the power Doppler is within \qty{9}{\dB} of its 99.5th percentile and the speed exceeds \qty{0.1}{\meter\per\second}.
    (e)~\texttt{twente-vortexflow}, \qty{120}{\volt} pump: fluctuating Doppler velocity along the wake of the cylinder against time; the diagonal stripes are vortices carried downstream.
    (f)~\texttt{twente-microbubblesim}: B-mode of the short \qty{2.5}{\MHz} pulse, true bubble positions circled, solid within \qty{\pm 2}{\mm} of the imaging plane and faint beyond.}
    \label{fig:flow}
\end{figure}

\paragraph{3D Acquisitions}
A number of contributions enable positioning their frames in 3D, as shown in Fig.~\ref{fig:slices3d}.
In \texttt{mosaic-nuevosono}, an \ac{IVUS} catheter spins a single element inside the vessel and is pulled back along it, so each rotation images a cross-section centered on the transducer, and the cross-sections stack along a \qty{67}{\mm} pullback (a).
In \texttt{tumunich}, a robot translates a linear array across its elevation, stacking parallel planes over \qty{50}{\mm} of a thyroid phantom (b).
In \texttt{strasbourg-basel}, a robot instead follows a fractured femur phantom along a curved \qty{276}{\mm} path, so the image planes turn with the bone (c).
In \texttt{wpi}, the probe stays in place and the array is rotated by \ang{181} about its own axial axis, producing a fan of planes that share one axis, as used for elevational synthetic-aperture focusing (d).
In \texttt{us4us}, the subject moves rather than the probe, stepping a forearm through a ring array to obtain stacked tomographic slices (e).
The frame spacing alone spans more than two orders of magnitude, from \qty{0.04}{\mm} between frames (\texttt{tumunich}) through \qty{0.6}{\mm} (\texttt{mosaic-nuevosono}) and \qty{1.75}{\mm} (\texttt{strasbourg-basel}) to \qty{9}{\mm} between slices (\texttt{us4us}).

\subsubsection{Blood flow / ultrasound localization microscopy}
Six of the \OpenHRFContributors contributions target flow estimation, across simulation, phantom and \emph{in-vivo}, in 2D and 3D, see Fig.~\ref{fig:flow}.
\texttt{ulmshare} contains 99 contrast-enhanced acquisitions of the mouse brain through the intact skull, from 61 animals, enough to develop and train \ac{ULM} methods (Fig.~\ref{fig:flow}). From \qty{12.8}{\second} of one acquisition, power Doppler resolves the vasculature (a) and \ac{ULM} traces \num{3016} microbubble tracks (b).
\texttt{resolvestroke-clinical} and \texttt{waterloo-carotid} store their own flow maps alongside the raw channel data, so new methods can be compared with theirs on identical input: 3D transcranial perfusion of the human brain (c), and vector flow in 93 carotid acquisitions from eight participants (d).
Two Twente contributions add a flow phantom whose vortex street is visible in the Doppler velocity (\texttt{twente-vortexflow}, e) and simulated microbubbles with every true position stored (\texttt{twente-microbubblesim}, f).
\begin{figure}[!t]
    \centering
    \includegraphics[width=\linewidth]{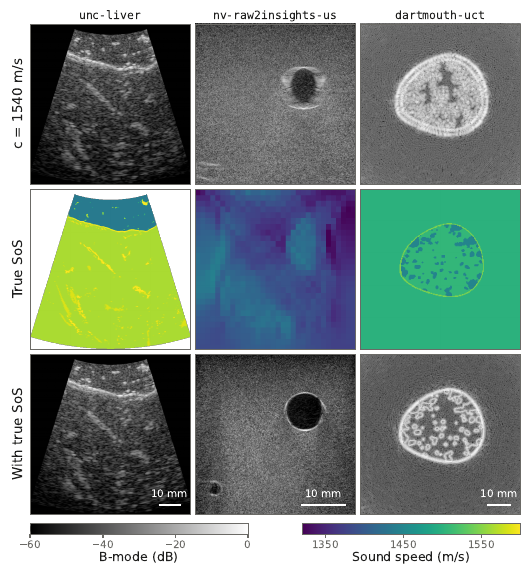}
    \caption{\acs{SoS} correction on three of the four simulations that store their \acs{SoS} map. Columns: \texttt{unc-liver} abdominal wall over liver (curvilinear array), \texttt{nv-raw2insights-us} phantom with an aberrating layer and cysts (linear array), and \texttt{dartmouth-uct} breast in a ring array.
    Top: \acs{DAS} at a uniform \qty{1540}{\meter\per\second}. Middle: stored \acs{SoS}. Bottom: the same beamformer with straight-ray delays through the true map.}
    \label{fig:sos}
\end{figure}

\subsubsection{Speed-of-sound estimation / aberration correction}
Nine of the \OpenHRFContributors contributions name \ac{SoS} estimation or aberration correction as an intended use.
Four simulations store the full \ac{SoS} map that generated the data: a Fullwave abdominal wall over liver (\texttt{unc-liver}), a phantom with an aberrating layer and cysts (\texttt{nv-raw2insights-us}), and two \ac{USCT} sets, a breast in a ring array (\texttt{dartmouth-uct}) and a limited-view prostate (\texttt{unc-openpros}).
Two others provide measured sound speeds: \texttt{stanford-murine} images phantoms beneath \qtyrange{10}{15}{\mm} layers of pork or chicken, whose \ac{SoS} was measured by through-transmission, and \texttt{waterloo-muscle} records a through-transmission \ac{SoS} of the calf alongside its \emph{in-vivo} muscle acquisitions.
The remaining three, \emph{in-vivo} breast lesions with biopsy-based labels (\texttt{kaist-snubh-barreleye}) and thyroid (\texttt{weizmann-sampl}), and a robot-tracked fractured-femur phantom (\texttt{strasbourg-basel}), do not include any numerical reference.

Fig.~\ref{fig:sos} shows how the stored maps can be used in full-synthetic-aperture simulations to update reconstruction assumptions.
Each dataset is beamformed twice with the same straight-ray delay model: once through a uniform \qty{1540}{\meter\per\second} medium, and once through its true map.
In the \texttt{nv-raw2insights-us} phantom, the medium is \qtyrange{100}{200}{\meter\per\second} slower than assumed, which smears the cyst into an ellipse at \qty{1540}{\meter\per\second}.
In \texttt{dartmouth-uct}, the ring array sees the breast through water, so assuming \qty{1540}{\meter\per\second} blurs the skin and merges the inclusions.
The \texttt{unc-liver} abdominal wall changes little: the slow fat layer ($\approx$\qty{1440}{\meter\per\second}) and the faster tissue below ($\approx$\qty{1565}{\meter\per\second}) average out close to \qty{1540}{\meter\per\second} along each path.

\subsection{Dataset Composition}
\label{sec:composition}
\begin{figure*}[!b]
    \centering
    \includegraphics[width=1\linewidth]{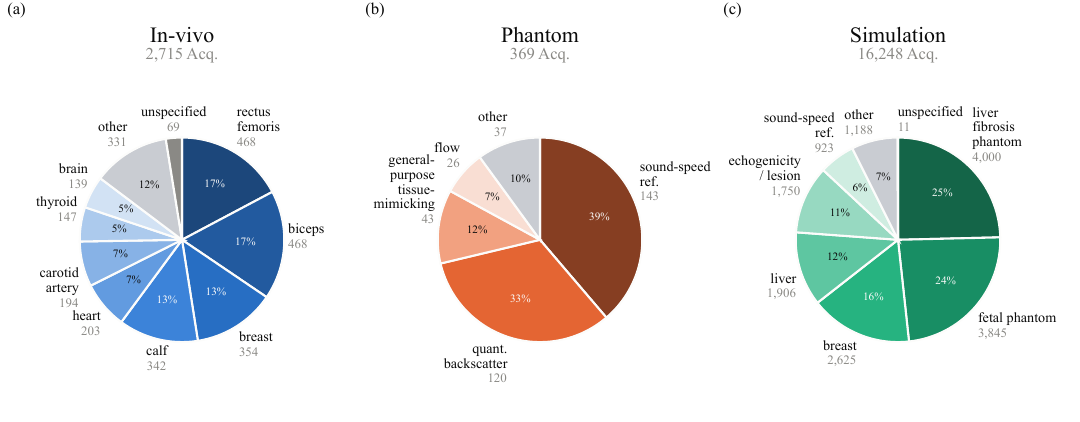}
    \caption{All \OpenHRFAcquisitions acquisitions divided by medium into (a) \emph{In-vivo}, (b) \textit{Phantom}, and (c) \textit{Simulation}, and further divided into anatomical / phantom subtypes.}
    \label{fig:piecharts}
\end{figure*}

\begin{figure*}
    \centering
    \includegraphics[width=\linewidth]{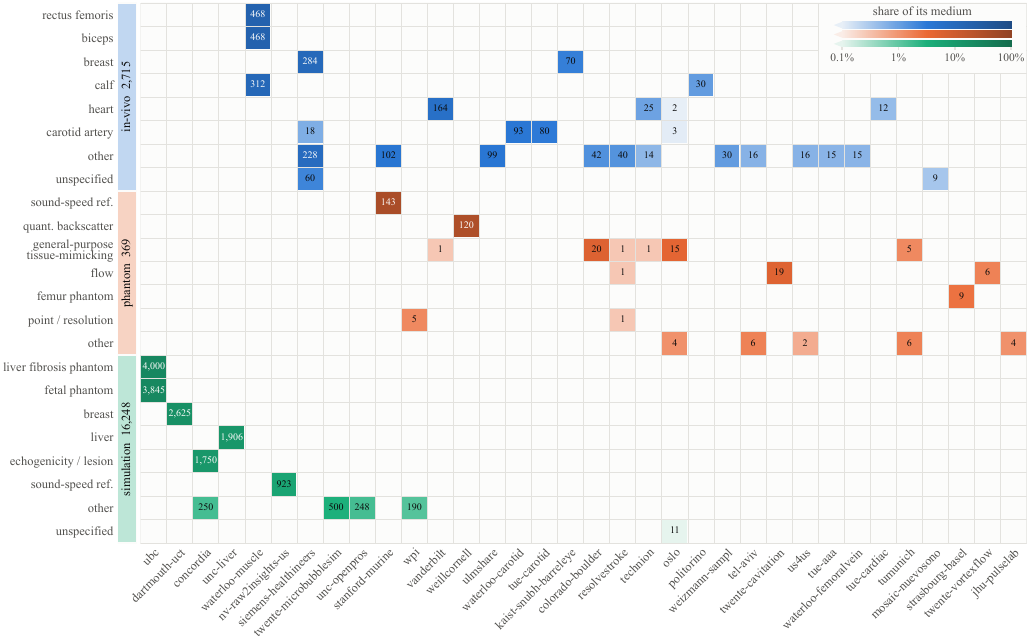}
    \caption{A matrix indicating which anatomy / phantom subtypes are contained by each contribution to \OpenHRF. The brightness of each color indicates how much of a medium's share is contained by that contribution. Additionally, the number of acquisitions is displayed in each square.}
    \label{fig:subtypes_to_datasets}
\end{figure*}

Version \texttt{v1.1} of \OpenHRF comprises \OpenHRFContributors contributions, consisting of \OpenHRFAcquisitions acquisitions, \OpenHRFFrames frames, and \OpenHRFTransmitEvents transmit events. The acquisitions cover a range of imaging media, including both \emph{in-vivo} and phantom targets, as well as a large number of \emph{in-silico} simulations. Fig.~\ref{fig:piecharts} categorizes these acquisitions in terms of imaging medium and corresponding subtypes. Fig.~\ref{fig:piecharts} shows that each medium contains a wide range of subtypes, encompassing many of the common ultrasound imaging targets. Fig.~\ref{fig:subtypes_to_datasets} maps this categorization onto the contributions, identifying, for example, that 4 contributions image the carotid artery.

\begin{figure}
    \centering
    \includegraphics[width=0.95\linewidth]{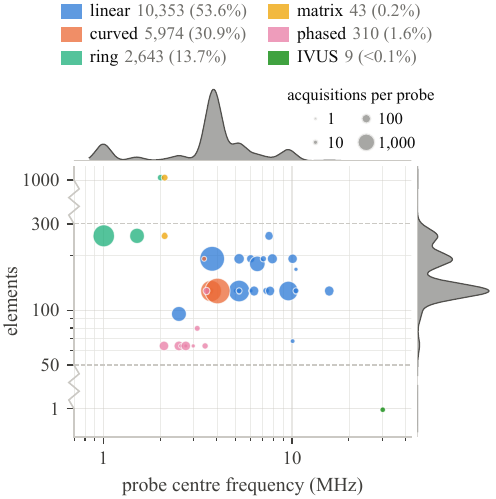}
    \caption{Acquisitions divided by probe type, with probe center frequencies plotted against element counts.}
    \label{fig:probes}
\end{figure}

\begin{figure}
    \centering
    \includegraphics[width=\linewidth]{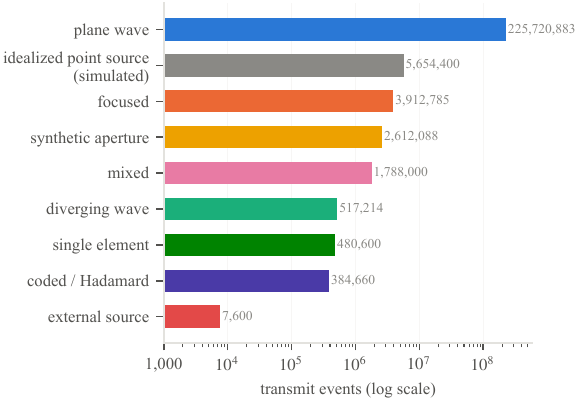}
    \caption{Transmit events sorted by transmit scheme. \textit{Mixed} refers to acquisitions that interleave multiple transmit schemes in one track; \textit{single element} refers to mechanically scanned single-element transmits; \textit{external source} marks tracks in which a separate probe is used for insonification.}
    \label{fig:transmit_schemes}
\end{figure}

The contributions also cover a wide variety of probe types and transmit schemes, illustrated by Figs.~\ref{fig:probes} and \ref{fig:transmit_schemes}, respectively. In terms of probe type, the contributions contain linear, curved, ring, phased-array, and \ac{IVUS} probes, with a range of transducer element counts and center frequencies. Common transmit schemes include plane wave, synthetic aperture, focused, diverging wave, coded, and \textit{mixed}, which refers to cases in which multiple transmit scheme types are used in a single track.

\subsection{Lag-One Coherence (LOC)}
\begin{figure}[htp]
    \centering
    \includegraphics[width=\linewidth]{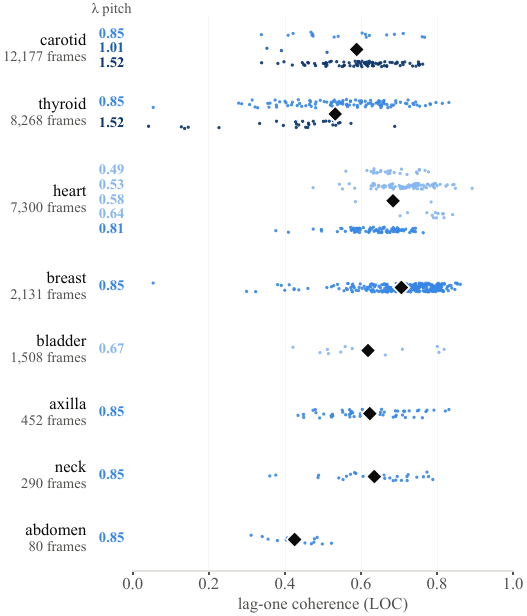}
    \caption{Lag-one coherence per anatomy across the focused \emph{in-vivo} datasets. Every dot denotes an acquisition available in \OpenHRF, the black diamond shows the mean for a given anatomy. Transducer pitch is expressed as a fraction of transmit wavelength $\lambda$ and separates the acquisitions vertically per anatomy.}
    \label{fig:loc}
\end{figure}

The \ac{LOC} is the normalized cross-correlation between the delayed signals of neighboring array elements~\cite{longLagOneCoherence2018}. This can give a local, target-free estimate of the achievable contrast and \ac{CNR}. With \OpenHRF, we supply data with a large \ac{LOC} spread, which allows evaluating methods under realistic noise and clutter conditions. We plot the \ac{LOC} for a subset of the collected \emph{in-vivo} anatomies in Fig.~\ref{fig:loc}, ordered by the number of frames. Every dot in the plot denotes an acquisition that was done with full-aperture focused imaging, single-element transmits, or line-scanning, with at least three frames available. The data comes from the \texttt{siemens-healthineers}, \texttt{tue-carotid}, \texttt{oslo-carotid}, \texttt{weizmann-sampl}, \texttt{vanderbilt} (both fundamental and harmonic), \texttt{technion-cardiac}, \texttt{tue-cardiac}, \texttt{oslo-cardiac} and \texttt{technion-bladder} contributions.

\section{Discussion}
Raw channel data is difficult and expensive to acquire. It requires research scanners that expose raw channel data, which are costly and not available in every lab. In addition, raw channel data is often considered proprietary, as it reveals details of the transmit sequences and system design. As a result, research on raw channel data has largely been confined to groups with their own acquisition setup, with methods developed and evaluated on in-house data rather than across institutions. \OpenHRF offers a shared, permissively licensed alternative, contributed by both academic and industrial partners.

The contents of \OpenHRF may be broken down in a number of ways, providing different views on its composition. For example, in terms of acquisitions, the \textit{simulation} medium dominates, covering \qty{84}{\percent} of all acquisitions. This could be explained by the relative ease of generating novel simulated subjects. If we consider the number of frames, however, \emph{in-vivo} dominates, with \qty{98}{\percent} of all frames being \emph{in-vivo}, and \qty{85}{\percent} of those \emph{in-vivo} frames coming from \texttt{ulmshare}, which contains long acquisitions imaged at a high frame rate. Consecutive frames from such acquisitions are highly correlated, so frame counts overstate how much independent information a contribution adds; for anatomical and physiological diversity, the number of distinct subjects, animals, or phantoms is a better proxy for dataset size. The unevenness in representation of different imaging media or anatomical types therefore depends heavily on the level of abstraction we use to count the data: \textit{acquisition}, \textit{frame}, or \textit{transmit event}, and should be taken into account when sampling across contributions for benchmarking or model training.

\OpenHRF is designed to support a number of tasks, as presented in Table \ref{tab:dataset-tasks}. Table \ref{tab:task-counts} shows the \textit{task coverage}, measured by the number of contributions supporting each task, and each task category. There is broad coverage of most of the task categories, with \textit{Generalized reconstruction / beamforming} being supported by most of the contributions (28), and all other tasks being covered by at least 5 contributions. The exception is the \textit{Therapy} category, containing only one task, \textit{Microbubble cavitation detection}, which is covered only by the \texttt{twente-cavitation} dataset. This insight can serve to guide future data collection, to improve the representation and coverage of therapeutic tasks.

\begin{table}[!ht]
  \centering
  \caption{Number of \OpenHRF contributions declaring each task, grouped by task category.}
  \label{tab:task-counts}
  \setlength{\tabcolsep}{4pt}
  \begin{tabularx}{\columnwidth}{@{}>{\raggedright\arraybackslash}p{2.0cm}>{\raggedright\arraybackslash}Xr@{}}
    \toprule
    Category & Task & Contributions \\
    \midrule
    Probe motion            & 3D / tracked reconstruction                  &  6 \\
    \midrule
    Tissue motion           & Motion estimation / tracking                 & 10 \\
                            & Blood flow / ultrasound localization microscopy & 6 \\
    \midrule
    Inaccurate assumptions  & Generalized reconstruction / beamforming     & 28 \\
                            & Aberration correction                        &  9\textsuperscript{\textdagger} \\
                            & Clutter / reverberation suppression          &  6 \\
                            & Image-quality assessment                     &  8 \\
    \midrule
    Tissue characterization & Speed-of-sound estimation                    &  9\textsuperscript{\textdagger} \\
                            & Quantitative ultrasound                      &  5 \\
                            & Segmentation / interpretation                &  7 \\
    \midrule
    Therapy                 & Microbubble cavitation detection             &  1 \\
    \bottomrule
  \end{tabularx}
  \par\smallskip\raggedright\footnotesize\textsuperscript{\textdagger}Aberration correction and speed-of-sound
  estimation are a single annotation category; the same 9 contributions are counted under both.
\end{table}

A key capability enabled by \OpenHRF is the evaluation of task performance across scanners, probes, and institutions. This makes it possible to study how they generalize and gives a common basis for benchmarking. In addition to task-specific methods, \OpenHRF opens new possibilities in developing methodology for imaging efficiency and quality enhancement that generalizes across tasks. For example, \textit{Compressed Sensing} \cite{yousufi2019application, huijben2020learning} may be used to recover useful information from fewer measurements, or select transmissions according to the information needed for reconstruction or a downstream task. An interesting avenue for research in this direction is to compare the information gain or quality improvement offered by different transmit schemes, such as focused, plane-wave, diverging-wave, and coded transmissions. This is supported by a number of contributions (e.g., \texttt{jhu-pulselab}, \texttt{oslo}, \texttt{stanford-murine}, \texttt{tue-cardiac}, \texttt{tue-carotid}, \texttt{twente-microbubblesim}), which image the same target using different transmit schemes, stored either as parallel \zea tracks or as separate acquisitions.

The large volume and variety of raw channel data provided by \OpenHRF opens up new possibilities in developing data-driven solutions that operate on raw channel data. Interesting opportunities lie in both supervised and unsupervised learning tasks. On the supervised side, training regression and classification models directly on raw channel data could circumvent errors and information loss introduced by the beamforming step \cite{abbey2010optimal}. Research efforts in this direction could explore methods to standardize data across anatomies, probes, etc., as well as neural network architectures that are parameter-efficient and extract useful features. Unsupervised methods, e.g., self-supervised learning, could be used to train generative models of raw channel data, capturing its structure and variation. Potential applications of raw channel data generation include image quality enhancement (e.g., dehazing \cite{stevensDehazingUltrasoundUsing2024}) and raw channel data compression~\cite{federiciVariationalUltrasoundChannel2026}. Another application of such a model is \textit{cognitive ultrasound}~\cite{vanslounActiveInferenceDeep2024,vannieropPatientAdaptiveEchocardiographyUsing2026}, which harnesses data-driven generative models to achieve information-seeking, patient-adaptive ultrasound acquisitions, and would benefit from a generative prior over the raw channel data itself rather than over the reconstructed image~\cite{stevensDeepGenerativeModels2025}.
A natural next step is a foundation model for raw channel data. Such a model could be pretrained with self-supervised objectives on the full corpus and conditioned on the probe geometry and transmit parameters stored in every file, allowing one model to handle different probes and transmit schemes.

The dataset also lowers the barrier for researchers outside the ultrasound community. Each acquisition is stored as a self-contained file with its reconstruction pipeline and can be streamed from the Hugging Face Hub, which makes working with raw channel data considerably easier without extensive ultrasound expertise. This makes it an interesting and challenging benchmark for the computational imaging and machine learning communities. We see \OpenHRF as a starting point rather than a finished resource. The format and submission process are designed to accept new contributions, and we invite the community to extend the corpus, define benchmarks on it, and share the models they train on it.

\section{Conclusion}
We present \OpenHRF: the largest openly licensed collection of raw medical ultrasound data to date. The dataset is contributed by \OpenHRFInstitutions institutions from both academia and industry and comprises approximately \OpenHRFSize of compressed raw channel data from \OpenHRFAcquisitions ultrasound scans. It covers a diverse set of tasks, ranging from general reconstruction to motion and blood flow estimation, as well as tracked ultrasound. Throughout, we place a strong focus on accessibility for researchers who want to work with this kind of data, as reflected in the permissive license and the open-source ecosystem in which the dataset is hosted. We hope that \OpenHRF will enable researchers to develop methods that require raw channel data at scale, and unlock rigorous evaluation and discovery in this traditionally restricted research space. We hope to make ultrasound more accessible to researchers and build bridges to other communities, e.g., machine learning and computational imaging communities. One direct and promising direction is the development of an ultrasound foundation model trained on raw channel data and conditioned on the provided metadata. Ultimately, we hope \OpenHRF serves as a shared resource for the community to build upon, moving ultrasound research beyond the beamformed image and towards methods that exploit the full information content of the raw signal.

\section*{Acknowledgment}
\textbf{Tristan S.W. Stevens} Data Curation, File Format Design, Software, Writing -- Original Draft. \textbf{Simon W. Penninga} Data Curation, Writing -- Original Draft, Visualization, Marketing. \textbf{Oisín Nolan} File Format Design, Software, Writing -- Original Draft, Visualization. \textbf{Wessel L. van Nierop} File Format Design, Software, Writing -- Original Draft.
\textbf{Felix Duelmer} Data Curation, Software, Writing -- Original Draft. \textbf{Ben N. Frey} Data Curation.
\textbf{Louis van Harten} Website Creation, Visualization.
\textbf{Ruud J.G. van Sloun} Conceptualization, Supervision. \textbf{Jeremy Dahl} Conceptualization, Supervision. \textbf{Walter Simson} Conceptualization, Supervision, Data Curation, Writing -- Review \& Editing.

All other authors are dataset contributors from the following institutions:
\textit{
Eindhoven University of Technology, Technical University of Munich, Stanford University, Mosaic Intelligence, Technion – Israel Institute of Technology, Concordia University, Dartmouth College, IHU Strasbourg, ImFusion, Johns Hopkins University, KAIST, Barreleye Inc., National Institutes of Health, NuevoSono, Politecnico di Torino, Ruhr University Bochum, Polytechnique Montréal, Mila, Resolve Stroke, Seoul National University Bundang Hospital, Siemens Healthineers, School of Biomedical Engineering Tel-Aviv University, The University of British Columbia, University of Basel, University of Colorado Boulder, University of North Carolina at Chapel Hill, North Carolina State University, Iowa State University, QT Imaging Inc., University of Oslo, University of Rochester, University of Twente, University of Waterloo, us4us Ltd., Vanderbilt University, Weill Cornell Medicine, Weizmann Institute of Science, Worcester Polytechnic Institute, NVIDIA%
}.

\printbibliography

\end{document}